\documentclass[11pt]{article}
\usepackage[dvipdfmx]{graphicx}
\usepackage{url}
\usepackage{typearea}
\typearea{20}
\usepackage{hyperref}
\usepackage{amsmath}

\providecommand{\keywords}[1]
{
  \small	
  \textbf{\textit{Keywords---}} #1
}

\begin{document}

\title{Empirical Evaluation and Scalability Analysis of Proof of Team Sprint (PoTS): Reward Fairness, Energy Efficiency, and System Stability}
\author{Naoki Yonezawa \thanks{\texttt{n.yonezawa@thu.ac.jp}. Faculty of Humanities and Social Sciences, Teikyo Heisei University, Japan.}}
\date{}

\maketitle

\begin{abstract} This paper presents a comprehensive empirical evaluation of the Proof of Team Sprint (PoTS) consensus algorithm, focusing on its impact on reward fairness, energy efficiency, system stability, and scalability. To assess these factors, we conducted large-scale simulations comparing PoTS with conventional Proof of Work (PoW) under various computational environments and team configurations. Our results demonstrate that PoTS significantly mitigates reward disparities caused by heterogeneous computational capabilities. In PoW, the highest-performing node secured the top ranking in 100 out of 100 trials, leading to extreme centralization. In contrast, under PoTS, the dominance of high-performance nodes was reduced, with the strongest node ranking first only 54 times out of 100, ensuring a fairer distribution of rewards among participants. The skewness and kurtosis of the reward distribution decreased as team size increased, confirming that PoTS fosters a more equitable allocation of mining rewards. Energy efficiency analysis revealed that PoTS exhibits a near $1/N$ scaling trend in computational workload, where $N$ is the team size. Total active computation time in PoTS (team size 64) was reduced by a factor of 64 compared to PoW, demonstrating a drastic reduction in energy consumption while maintaining the same level of security and consensus reliability. Furthermore, system stability and scalability evaluations confirmed that PoTS maintains statistical consistency across multiple simulation runs, ensuring robust performance regardless of network size. The correlation between participant performance and reward allocation increased with team size, peaking at 0.90 when $N=16$, before stabilizing, highlighting the controlled balance between computational power and fairness. These findings suggest that PoTS is a viable alternative to traditional PoW, offering a more decentralized, fair, and energy-efficient approach to achieving blockchain consensus. This study provides empirical validation for PoTS's ability to improve the sustainability and fairness of decentralized networks, positioning it as a promising consensus mechanism for future blockchain applications. \end{abstract}


\section{Introduction}
\subsection{Motivation}

The increasing adoption of blockchain technology has been accompanied by concerns regarding its energy consumption and centralization risks. The most widely used consensus mechanism, Proof of Work (PoW), ensures security through computational competition, requiring miners to solve complex cryptographic puzzles to validate transactions. However, this approach results in significant energy inefficiency, with the total computational power of the network being orders of magnitude higher than what is theoretically required to maintain security. The competitive nature of PoW further exacerbates centralization, as large-scale mining operations with access to specialized hardware dominate the network, reducing decentralization.

To address these issues, \textit{Proof of Team Sprint} (PoTS)~\cite{yonezawa2024pots} has been proposed as a novel consensus mechanism that shifts from an individual competition model to a collaborative, team-based approach. PoTS organizes participants into groups, requiring them to work together to generate blocks in a sequential manner. This collaborative structure significantly reduces redundant computations and distributes computational effort across team members, leading to enhanced energy efficiency. Moreover, PoTS mitigates centralization by ensuring that rewards are shared within teams, promoting fairness and reducing the dominance of high-performance nodes.

While the theoretical benefits of PoTS have been outlined in previous work, a rigorous empirical evaluation is needed to validate its effectiveness. Specifically, it is crucial to quantify how PoTS impacts reward distribution fairness, energy efficiency, and system stability across different team sizes. This study aims to provide a comprehensive empirical assessment of PoTS, comparing it with PoW through large-scale simulations.

\subsection{Contributions}

This paper presents a detailed empirical analysis of PoTS, comparing it against PoW in terms of fairness, energy efficiency, and decentralization. Our key contributions are as follows:

\begin{itemize}
    \item \textbf{Empirical evaluation of reward distribution fairness:} We analyze how PoTS redistributes rewards compared to PoW, demonstrating its ability to mitigate centralization and enhance fairness among participants.
    \item \textbf{Quantitative analysis of energy efficiency:} We measure the total active computation time under PoTS and PoW, illustrating how PoTS significantly reduces energy consumption.
    \item \textbf{System stability and scalability:} We investigate the impact of team size on PoTS performance, examining how increasing the number of participants in a team affects reward distribution and computational workload.
    \item \textbf{Impact of high-performance nodes:} We assess the influence of computational heterogeneity by introducing a high-performance node and evaluating its dominance under PoTS and PoW.
\end{itemize}

The remainder of this paper is structured as follows: Section~\ref{sec:background} provides an overview of PoW and PoTS, along with related experimental studies. Section~\ref{sec:methodology} details our simulation setup and evaluation metrics. Section~\ref{sec:results} presents our findings on reward distribution, energy efficiency, and system scalability. Section~\ref{sec:discussion} discusses the implications of our results, and Section~\ref{sec:conclusion} concludes the study with directions for future work.

\section{Background and Related Work}
\label{sec:background}

\subsection{Overview of Proof of Work (PoW)}

Proof of Work (PoW) is the original consensus mechanism used in Bitcoin~\cite{nakamoto2008peer} and remains widely deployed in blockchain networks. In PoW, miners compete to solve cryptographic puzzles, with the first to find a valid solution earning the right to add a new block to the blockchain. This mechanism ensures network security by making attacks computationally expensive. However, PoW suffers from two major issues:

\begin{itemize}
    \item \textbf{High energy consumption:} The mining process requires enormous computational power, leading to excessive energy consumption. Recent estimates as of September 2024 suggest that the energy consumption of the Bitcoin network alone rivals that of entire countries, such as Poland or Malaysia.
    \item \textbf{Centralization risks:} The reliance on specialized mining hardware (ASICs) favors large-scale operations, leading to a concentration of mining power among a few entities.
\end{itemize}

Several alternative consensus mechanisms have been proposed to address these issues, including Proof of Stake (PoS), Delegated Proof of Stake (DPoS), and energy-efficient PoW variants. These approaches aim to reduce energy consumption while maintaining security but often introduce new trade-offs, such as potential wealth concentration in PoS-based systems.

\subsection{Proof of Team Sprint (PoTS)}

Proof of Team Sprint (PoTS)~\cite{yonezawa2024pots} is a novel consensus mechanism designed to mitigate the inefficiencies of Proof of Work (PoW) while maintaining decentralization. Unlike PoW, where miners compete individually to solve cryptographic puzzles, PoTS organizes participants into teams, enabling a collaborative approach to block generation. This team-based strategy reduces redundant computations, ensuring that energy is used more efficiently across the network.

\begin{figure}[t]
\begin{center}
\includegraphics[scale=0.5]{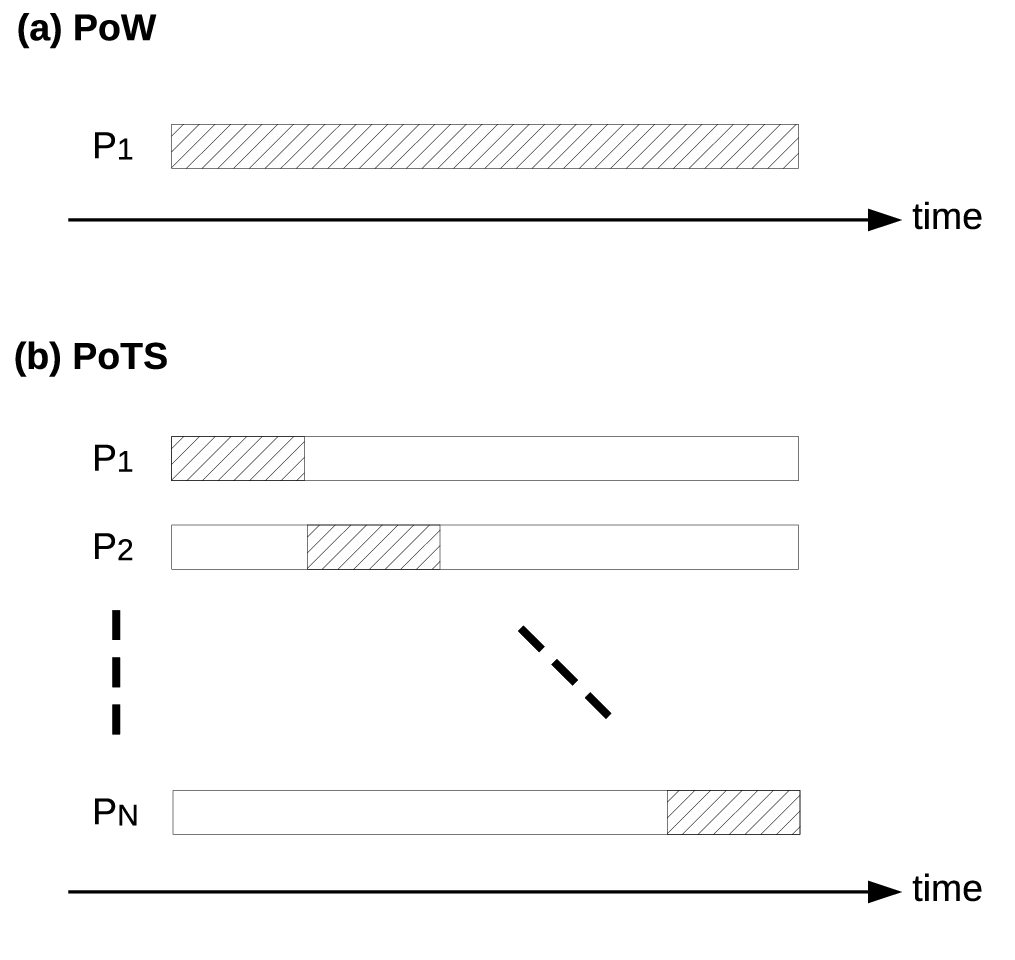}
\caption{Comparison of Block Generation Process in PoW and PoTS. Adapted from~\cite{yonezawa2024pots}.}
\label{fig:pow-pots-time-diagram}
\end{center}
\end{figure}

In PoTS, each team is assigned a fixed computation time per round, and participants take turns contributing to block creation in a sequential manner, as illustrated in Figure~\ref{fig:pow-pots-time-diagram}. This structured approach minimizes energy wastage and enhances fairness by distributing workload among multiple participants. Unlike PoW, where only the fastest miner receives a reward, PoTS ensures that all team members who contribute to block creation share the rewards.

Key features and advantages of PoTS include:

\begin{itemize}
    \item \textbf{Energy-efficient computation:} The total computational effort per round remains constant regardless of team size. Since work is distributed among participants, the energy consumption per participant is theoretically reduced by a factor of $1/N$, where $N$ is the team size. This results in significant energy savings compared to PoW, where all miners independently expend energy competing to solve the same problem.
    \item \textbf{Fair reward distribution:} Unlike PoW, where high-performance miners with specialized hardware dominate the mining process, PoTS ensures equitable reward distribution. By structuring mining as a cooperative effort, PoTS prevents monopolization of rewards by a few dominant participants, fostering a more inclusive blockchain network.
    \item \textbf{Mitigation of high-performance node dominance:} In PoW, miners with superior computational resources have a significant advantage in block generation, leading to centralization. PoTS mitigates this by requiring team-based sequential computation, ensuring that no single high-performance node can unilaterally dominate the consensus process. This makes PoTS more resistant to miner centralization and strengthens the network's decentralization.
    \item \textbf{Scalability and adaptability:} PoTS is designed to scale efficiently with network growth. By dynamically adjusting team sizes and workload distribution, PoTS can maintain high transaction throughput without the energy-intensive scaling issues associated with PoW.
    \item \textbf{Robust security model:} The sequential nature of PoTS computations ensures that each participant's contribution is verifiable, reducing the risk of fraudulent activity. Since team composition is randomized in each round, collusion attacks are significantly harder to execute compared to PoW-based mining pools, where miners can form alliances to control block generation.
\end{itemize}

PoTS represents a significant departure from traditional competition-based mining models, introducing a structured and cooperative approach to blockchain consensus. By reducing computational redundancy, lowering energy consumption, and promoting fair participation, PoTS offers a sustainable alternative to PoW. In subsequent sections, we further analyze PoTS's performance, particularly in terms of energy efficiency, system stability, and resilience to high-performance node dominance.

\subsection{Related Experimental Studies}

Several studies have examined alternative consensus mechanisms and their trade-offs in terms of security, scalability, and energy efficiency. Previous research on PoS~\cite{bano2019sok} and DPoS~\cite{saad2021comparison} has highlighted their ability to reduce energy consumption while introducing potential risks related to wealth centralization. Similarly, Byzantine Fault Tolerance (BFT)-based mechanisms such as PBFT~\cite{castro1999practical} have been explored for permissioned blockchains but suffer from scalability limitations due to high communication overhead.

Experimental evaluations of energy-efficient consensus mechanisms have primarily focused on theoretical analyses or small-scale simulations. Studies such as~\cite{foytik2020blockchain} and~\cite{alharby2019blocksim} have introduced blockchain simulation frameworks to evaluate consensus algorithms under diverse network conditions. Additionally, research on simulation methodologies~\cite{paulavivcius2021systematic} provides insights into optimizing blockchain modeling, particularly for PoW and PoS systems.

Energy efficiency in consensus mechanisms has been a central focus of research. The Green-PoW framework~\cite{lasla2022green} proposes modifications to the PoW process, reducing computational redundancy and energy consumption. Similarly, research comparing PoW and PoS energy consumption~\cite{zhang2020evaluation} suggests that hybrid models can optimize energy use while maintaining network security. These studies provide valuable comparisons for PoTS's energy efficiency assessment.

The fairness of reward distribution is another key concern in blockchain consensus research. PoS-based systems often exhibit wealth accumulation effects~\cite{bandelli2021inequality}, leading to concerns about long-term decentralization. Simulation-based studies on PoS fairness~\cite{zhao2022nxt} and validator behavior~\cite{nguyen2019proof} indicate that even non-PoW mechanisms can exhibit centralization tendencies. PoTS addresses these concerns by ensuring that rewards are distributed among all team members, mitigating dominance by high-performance nodes.

In addition to fairness and energy efficiency, network security remains a critical issue. Research on PoW security~\cite{gervais2016security} has highlighted vulnerabilities such as selfish mining and double-spending attacks. Blockchain simulators~\cite{faria2019blocksim, wuthier2021proof} have been used to analyze network attack scenarios, including eclipse attacks and partitioning. PoTS's collaborative approach inherently reduces the risk of miner centralization, thus indirectly strengthening security against these threats.

Another aspect of experimental research involves scalability. Studies on permissioned blockchains~\cite{dabbagh2021survey} and consensus scalability~\cite{mardiansyah2022simblock} discuss the limitations of existing mechanisms under high transaction loads. PoTS's team-based structure offers a potential solution by allowing network participants to distribute workload dynamically, preventing bottlenecks seen in traditional PoW-based models.

Lastly, new simulation frameworks such as JABS~\cite{yajam2023jabs} have been developed to evaluate various consensus mechanisms under real-world conditions. These tools provide valuable methodological insights applicable to PoTS simulations, particularly for assessing consensus formation and network stability.

Our work builds upon these efforts by conducting large-scale simulations to empirically validate PoTS's ability to enhance fairness, improve energy efficiency, and maintain network stability. By integrating findings from these studies, we aim to position PoTS within the broader landscape of sustainable and decentralized blockchain consensus mechanisms.

\section{Experimental Methodology}
\label{sec:methodology}
\subsection{Simulation Setup}

\subsubsection{Simulation Settings}
In our simulation configuration, we modeled a mining system for the fictional cryptocurrency ``SprintCoin'' with a total of $n = 1,600$ participants. Each participant was assigned a performance factor drawn from a uniform distribution between 0.8 and 1.5. To specifically examine the impact of a high-performance node, we manually set the performance factor of a single participant (ID 1000) to 2.5, maintaining the remaining participants within the initial range.

Throughout the simulation, participants were grouped into teams of fixed sizes, denoted as $N$, and the number of teams per round was $M = n/N$. The teams were randomly reassembled at the start of each new round. A total of 1,600 rounds were conducted for each simulation scenario. The specific team sizes tested and the corresponding number of teams formed per round are listed in Table~\ref{tab:team_sizes}.

\begin{table}[h]
\centering
\begin{tabular}{|c|c|}
\hline
\textbf{Team Size} ($N$) & \textbf{Number of Teams} ($M$) \\
\hline
1 (equivalent to PoW) & 1600 \\
2 & 800 \\
4 & 400 \\
8 & 200 \\
16 & 100 \\
32 & 50 \\
64 & 25 \\
\hline
\end{tabular}
\caption{Team sizes ($N$) and corresponding number of teams ($M$) in simulation runs.}
\label{tab:team_sizes}
\end{table}

Each team was assigned a fixed base computation time of 600 seconds per round. Within each team, participants took turns processing, sequentially contributing to block generation. Since each team consisted of $N$ participants, the number of blocks produced per round also matched $N$, ensuring that all participants contributed equally within their respective teams.

Each participant was allocated an individual work time calculated as $600 / N$. However, the actual processing time of each participant varied due to a scaling factor sampled from a predefined range. The final mining time for each participant was computed as:

\begin{equation}
\text{simulated time} = \frac{\text{work time} \times \text{multiplier}}{\text{performance factor}}
\end{equation}

where the multiplier was a random value sampled from a specified range \([0.8, 1.2]\), introducing variability in block generation time.

To ensure consistency in workload, the total computation required per round remained constant. That is, the total amount of work was adjusted such that, regardless of team size, the cumulative computational effort per round was equivalent to 600 seconds of work by a single participant.

Each simulation round awarded a fixed reward of 10 units to the winning team, distributed equally among its $N$ members.

Given our specific focus on evaluating reward fairness, energy efficiency, and system stability, we assumed no network latency and that participants did not drop out mid-computation.

\subsubsection{Evaluation Metrics}

To assess the effectiveness of the PoTS consensus mechanism, we employ the following evaluation metrics, focusing on reward distribution, energy efficiency, and system fairness.

\begin{itemize}
  \item \textbf{Reward Distribution and Fairness:} We measure the distribution of rewards across participants by analyzing statistical properties such as mean, median, standard deviation, and percentiles. This helps evaluate whether PoTS mitigates reward centralization compared to PoW.
  
  \item \textbf{Total Active Computation Time:} We quantify the total computational effort expended per round under different team sizes. This metric provides insight into the energy efficiency of PoTS by examining how computation time scales with team size.
  
  \item \textbf{Correlation Between Reward and Performance Factor:} We compute correlation coefficients to assess the relationship between participant performance and received rewards. This helps determine whether PoTS reduces the direct impact of performance disparities compared to PoW.
  
  \item \textbf{Statistical Properties of Reward Distribution:} We examine skewness and kurtosis to analyze the shape of the reward distribution. These metrics help assess whether PoTS promotes a more equitable distribution compared to PoW.
\end{itemize}

These evaluation metrics provide a structured framework for analyzing PoTS's potential advantages in fairness, energy efficiency, and system stability.

\subsection{Experimental Scenarios}

To comprehensively evaluate the performance and fairness of PoTS, we designed several experimental scenarios. These scenarios allow us to assess the impact of team size, the presence of high-performance nodes, and the stability of the system under repeated trials.

\subsubsection{Independent Simulations for Different Team Sizes}
To examine how team size affects reward distribution and energy efficiency, we conducted independent simulations with varying team sizes, as summarized in Table~\ref{tab:team_sizes}. As previously mentioned, a team size of 1 corresponds to the PoW scenario, where each participant competes individually. For PoTS, we tested multiple team sizes (2, 4, 8, 16, 32, and 64) to analyze how workload distribution within teams influences system performance and fairness. Each configuration was evaluated under both PoTS and PoW conditions to enable direct comparison.

\subsubsection{Impact of High-Performance Nodes}
To evaluate the influence of computational heterogeneity, we introduced a single high-performance node into the simulation. As previously described, all participants had performance factors drawn from a uniform distribution between 0.8 and 1.5, except for one designated high-performance participant with a performance factor of 2.5. This setup allowed us to compare two conditions: one where all participants had similar performance factors and another where a single high-performance node was present. By analyzing these scenarios, we assessed whether PoTS mitigates the dominance of high-performance nodes compared to PoW.

\subsubsection{Reproducibility and Stability Experiments}
To ensure the robustness of our findings, each experimental scenario was repeated 100 times. This approach allows us to collect sufficient statistical data and evaluate the stability of the reward distribution, energy efficiency, and overall system behavior. By analyzing the variance and consistency of results across multiple runs, we can determine the reproducibility of PoTS's benefits under different conditions.

\section{Results}
\label{sec:results}

This section presents the results of our simulations, which were conducted over 100 independent runs for each scenario. The analysis focuses on reward distribution, energy efficiency, and system stability. Statistical summaries of key metrics are provided in the following subsections.

\subsection{Reward Distribution Analysis}

\subsubsection{PoTS vs. PoW Reward Statistics}

Table~\ref{tab:reward_distribution} presents the reward distribution statistics for PoW and PoTS across various team sizes. The results indicate a stark contrast in reward distribution between the two mechanisms. In PoW, the median reward is 0.0, suggesting that more than half of the participants receive no rewards at all. Additionally, the standard deviation is extremely high (54.81), with a maximum reward of 592.80, illustrating significant centralization of rewards among a few participants.

In contrast, PoTS exhibits a more balanced reward distribution. As team size increases, the median reward gradually approaches the mean (10.0), and the standard deviation decreases significantly. For example, at a team size of 64, the median reward reaches 10.05, and the standard deviation drops to 2.38. This confirms that PoTS mitigates reward disparity and promotes fairness by distributing rewards more evenly among participants.

Furthermore, these results demonstrate the scalability of PoTS. Unlike PoW, where the reward distribution remains highly skewed regardless of the number of participants, PoTS exhibits improved fairness as team size increases. The decreasing standard deviation and increasing median reward indicate that PoTS maintains stability and consistency across different scales, ensuring that the system remains robust even with larger participant pools. This scalability highlights PoTS's ability to provide an equitable reward mechanism without compromising efficiency.

\begin{table}[h]
\centering
\begin{tabular}{|l|c|c|c|c|c|c|c|}
\hline
\textbf{Scenario} & \textbf{Mean} & \textbf{Std. Dev.} & \textbf{Min} & \textbf{25th Pctl} & \textbf{Median} & \textbf{75th Pctl} & \textbf{Max} \\
\hline
PoW & 10.00 & 54.81 & 0.00 & 0.00 & 0.00 & 0.00 & 592.80 \\
PoTS (2) & 10.00 & 21.42 & 0.00 & 0.00 & 0.00 & 5.61 & 130.30 \\
PoTS (4) & 10.00 & 13.06 & 0.00 & 0.00 & 3.50 & 16.69 & 66.35 \\
PoTS (8) & 10.00 & 8.63 & 0.00 & 2.50 & 7.58 & 16.22 & 41.04 \\
PoTS (16) & 10.00 & 5.70 & 0.04 & 5.02 & 9.38 & 14.34 & 28.19 \\
PoTS (32) & 10.00 & 3.72 & 1.84 & 6.99 & 9.94 & 12.81 & 21.09 \\
PoTS (64) & 10.00 & 2.38 & 3.97 & 8.15 & 10.05 & 11.78 & 16.91 \\
\hline
\end{tabular}
\caption{Reward distribution statistics for PoW and PoTS with varying team sizes. }
\label{tab:reward_distribution}
\end{table}
 
 \subsubsection{Effect of High-Performance Node on Ranking}

To examine the impact of computational heterogeneity, we introduced a single high-performance node (performance factor 2.5) into the network. As shown in Table~\ref{tab:high_perf_ranking}, in PoW, this node consistently ranked first in all 100 simulation runs, indicating absolute dominance. However, under PoTS, the dominance was reduced, with the node achieving first place in only 54 out of 100 runs, and ranking below first place in 46 cases. This suggests that PoTS inherently distributes rewards more evenly, preventing disproportionate accumulation by a single high-performance node.

\begin{table}[h]
\centering
\begin{tabular}{|c|c|c|}
\hline
\textbf{Rank} & \textbf{PoTS (Count)} & \textbf{PoW (Count)} \\
\hline
1 & 54 & 100 \\
2 & 13 & 0 \\
3 & 4 & 0 \\
4 & 4 & 0 \\
5 & 3 & 0 \\
6 & 2 & 0 \\
7 & 2 & 0 \\
8 & 3 & 0 \\
9 & 2 & 0 \\
10 & 1 & 0 \\
11 or lower & 12 & 0 \\
\hline
\end{tabular}
\caption{Ranking frequency distribution of the high-performance node (ID 1000) in PoTS and PoW.}
\label{tab:high_perf_ranking}
\end{table}

\subsection{Energy Efficiency Evaluation}

Table~\ref{tab:total_computation_time} summarizes the total active computation time for PoW and PoTS. As expected, PoTS significantly reduces computational workload as team size increases. The total active time follows an approximately $1/N$ scaling trend, confirming that PoTS maintains a fixed total work per round while distributing the effort among team members. Since computation time correlates directly with energy consumption, these results demonstrate PoTS's superior energy efficiency over PoW.

\begin{table}[h]
\centering
\begin{tabular}{|l|r|}
\hline
\textbf{Scenario} & \textbf{Total Active Time} \\
 & \textbf{($10^6$ seconds)} \\
\hline
PoW & 1,378.89 \\
PoTS (2) & 689.91 \\
PoTS (4) & 344.59 \\
PoTS (8) & 172.43 \\
PoTS (16) & 86.28 \\
PoTS (32) & 43.10 \\
PoTS (64) & 21.53 \\
\hline
\end{tabular}
\caption{Total active computation time (in $10^6$ seconds) for PoW and PoTS across different team sizes.}
\label{tab:total_computation_time}
\end{table}

\subsection{System Stability and Scalability}

Ensuring the stability and scalability of a consensus mechanism is crucial for its long-term viability. This section examines how PoTS maintains statistical consistency across multiple runs and scales effectively as team sizes increase.

Table~\ref{tab:skewness_kurtosis} presents the skewness and kurtosis of the reward distribution for PoW and PoTS. In PoW, skewness (6.85) and kurtosis (51.47) indicate an extremely asymmetric distribution where rewards are heavily concentrated among a small subset of participants. This distribution is consistent with the dominance of high-performance nodes in PoW mining, leading to increased centralization.

In contrast, as team size increases in PoTS, both skewness and kurtosis decrease significantly, approaching values closer to a normal distribution. This trend demonstrates that PoTS stabilizes reward distribution by reducing extreme outliers. At a team size of 64, skewness (0.008) and kurtosis (-0.66) indicate a nearly symmetric reward distribution, ensuring that rewards are more evenly spread among participants. This statistical consistency confirms that PoTS effectively mitigates centralization risks and promotes long-term fairness in reward allocation.

Scalability is another critical aspect of PoTS's design. Unlike PoW, where increasing network participation leads to escalating energy consumption and competition, PoTS allows for controlled scaling. As the network grows, teams can dynamically adjust their size while maintaining a fixed total computational workload per round. This ensures that the network can accommodate an increasing number of participants without significantly impacting computational efficiency. Furthermore, the structured nature of team-based block generation prevents bottlenecks commonly seen in PoW systems, where higher competition leads to increased variance in block discovery times.

Overall, the results suggest that PoTS provides a more stable and scalable alternative to PoW by balancing workload distribution, reducing reward concentration, and allowing for seamless scalability without compromising fairness or energy efficiency.

\begin{table}[h]
\centering
\begin{tabular}{|l|r|r|}
\hline
\textbf{Scenario} & \textbf{Skewness} & \textbf{Kurtosis} \\
\hline
PoW & 6.846 & 51.47 \\
PoTS (2) & 2.529 & 6.19 \\
PoTS (4) & 1.414 & 1.28 \\
PoTS (8) & 0.792 & -0.22 \\
PoTS (16) & 0.404 & -0.67 \\
PoTS (32) & 0.167 & -0.72 \\
PoTS (64) & 0.008 & -0.66 \\
\hline
\end{tabular}
\caption{Skewness and kurtosis of reward distribution for PoW and PoTS.}
\label{tab:skewness_kurtosis}
\end{table}

\subsection{Correlation Analysis}
Table~\ref{tab:correlation} shows the correlation between reward and performance factor. In PoW, the correlation is relatively weak (0.31). One major reason for this is the competitive nature of PoW: with 1,600 participants and only 100 rounds, many participants—regardless of their performance factor—never receive any rewards. Even high-performance participants may fail to win due to the probabilistic nature of mining, leading to a relatively low correlation between individual performance and rewards.

In contrast, PoTS introduces a team-based structure where rewards are shared among team members. This structural difference has a direct impact on the observed correlation. Initially, as team sizes increase ($N=2, 4, 8, 16$), the probability of forming a team with high-performance participants rises, leading to a stronger relationship between performance and reward. This is because teams with at least one high-performance participant are more likely to win, and rewards are distributed among the team members. As a result, the correlation peaks around 0.90 at a team size of 16.

However, when team sizes become larger ($N=32, 64$), the composition of teams becomes more balanced. The likelihood of forming a team that consists solely of high-performance participants decreases, and teams include a mix of high- and low-performance participants. Consequently, the dominance of individual high-performance participants is diluted, leading to a slight decline in correlation (from 0.90 at $N=16$ to 0.85 at $N=64$). This trend suggests that while performance remains a factor in reward distribution, PoTS mitigates the extreme centralization effects seen in PoW by enforcing team-based mining.

These findings indicate that PoTS retains a performance-based reward structure while ensuring a more equitable distribution of rewards compared to PoW. Furthermore, by dynamically adjusting team sizes, PoTS balances efficiency, fairness, and decentralization in a scalable manner.
	
\begin{table}[h]
\centering
 \begin{tabular}{|l|c|}
\hline
\textbf{Scenario} & \textbf{Correlation} \\
\hline
PoW & 0.307 \\
PoTS (2) & 0.663 \\
PoTS (4) & 0.831 \\
PoTS (8) & 0.892 \\
PoTS (16) & 0.898 \\
PoTS (32) & 0.882 \\
PoTS (64) & 0.855 \\
\hline
\end{tabular}
\caption{Correlation between reward and performance factor in PoW and PoTS.}
\label{tab:correlation}
\end{table}
 
\section{Discussion}
\label{sec:discussion} 

\subsection{Implications for Decentralized Consensus Mechanisms}

The results of our experiments highlight the potential of PoTS to address key challenges in decentralized consensus mechanisms, particularly in mitigating centralization and improving fairness.

One of the fundamental criticisms of PoW is its tendency toward centralization, where participants with superior computational resources accumulate a disproportionate share of rewards. As shown in Table~\ref{tab:high_perf_ranking}, in a PoW environment, a high-performance participant consistently dominates, securing first place in all 100 simulations. In contrast, PoTS redistributes this advantage by incorporating team-based competition. The results indicate that the high-performance participant only secured first place in 54 out of 100 runs under PoTS, demonstrating that reward centralization is significantly reduced. By distributing rewards among team members, PoTS fosters a more equitable system where participants with lower computational power can still earn rewards.

Energy efficiency is another major advantage of PoTS. As demonstrated in Table~\ref{tab:total_computation_time}, the total active computation time in PoTS decreases proportionally to the team size. Since computational effort is shared among team members while maintaining a fixed total workload per round, PoTS exhibits an approximately $1/N$ scaling trend. This implies that PoTS can drastically reduce energy consumption while preserving network security and stability, addressing one of the major concerns of PoW-based blockchains.

Overall, these findings suggest that PoTS is a viable alternative to traditional PoW, as it offers a more balanced and energy-efficient approach to achieving consensus while mitigating the risk of excessive centralization.

\subsection{Limitations of the Simulation Study}

Although the simulation results provide valuable insights into the advantages of PoTS, several limitations should be acknowledged.

First, our simulation assumes a uniform performance factor distribution for the majority of participants (0.8 to 1.5), with a single high-performance node (2.5). While this setup allows for controlled comparisons, real-world blockchain networks exhibit more diverse computational capabilities and economic incentives. Future studies should explore the impact of more complex performance distributions.

Second, our model does not account for network latency or the possibility of participants dropping out mid-computation. In real-world deployments, communication delays and node failures could impact consensus stability. Future simulations should incorporate these variables to better approximate real-world conditions.

Lastly, our study assumes that participants always act honestly within their assigned teams. However, strategic behaviors such as collusion or free-riding could emerge in practical implementations. Analyzing the game-theoretic aspects of PoTS in adversarial settings remains an important avenue for future research.

\subsection{Future Work}

To further validate PoTS, several areas of future research should be explored.

First, experiments in real-world blockchain networks should be conducted to assess PoTS under realistic conditions. Deploying PoTS in a testnet environment could provide valuable empirical data on its robustness and scalability.

Second, the applicability of PoTS beyond traditional blockchain settings should be investigated. While our study focuses on mining-based scenarios, PoTS could be adapted to other consensus models, such as proof-of-stake (PoS) or hybrid mechanisms that combine elements of PoW and PoS.

Finally, further optimizations to PoTS should be explored, particularly in terms of dynamic team formation and adaptive workload distribution. Incorporating mechanisms to optimize team selection based on historical performance or trust metrics could further enhance the efficiency and fairness of the protocol.

By addressing these research directions, PoTS has the potential to become a practical and scalable consensus mechanism that balances fairness, security, and energy efficiency.

\section{Conclusion}
\label{sec:conclusion} 
\subsection{Summary of Findings}

This study conducted a comprehensive evaluation of the Proof of Team Sprint (PoTS) consensus mechanism, focusing on reward fairness, energy efficiency, and system stability. Through extensive simulations with varying team sizes, we compared PoTS with the traditional Proof of Work (PoW) mechanism and analyzed its impact on network dynamics.

The results confirmed that PoTS significantly mitigates reward centralization. In PoW, high-performance participants monopolized rewards, creating an uneven distribution. In contrast, PoTS distributed rewards more equitably among team members, reducing the dominance of individual high-performance nodes (Table~\ref{tab:high_perf_ranking}).

Additionally, PoTS demonstrated substantial energy efficiency gains. By distributing computational effort across teams while maintaining a fixed total workload per round, PoTS exhibited a nearly $1/N$ scaling trend in computation time (Table~\ref{tab:total_computation_time}). This suggests that PoTS could significantly reduce blockchain energy consumption without compromising security or consensus reliability.

The system stability analysis further validated PoTS's ability to reduce reward distribution skewness and kurtosis, ensuring a more balanced reward structure across different network conditions (Table~\ref{tab:skewness_kurtosis}). Moreover, the correlation between reward and participant performance factor increased with team size, confirming that PoTS maintains a controlled balance between computational power and reward fairness (Table~\ref{tab:correlation}).

Overall, these findings suggest that PoTS is a viable alternative to PoW, offering a fairer and more energy-efficient approach to decentralized consensus.

\subsection{Final Remarks and Future Directions}

The adoption of blockchain technology continues to expand, driving demand for consensus mechanisms that prioritize fairness and efficiency. PoTS addresses these challenges by integrating team-based collaboration into mining operations, mitigating excessive centralization while preserving computational security.

Despite the promising results, further research is needed to validate PoTS in real-world blockchain implementations. Future studies should focus on deploying PoTS in test networks to assess its practical feasibility and potential challenges. Additionally, extending PoTS to hybrid consensus models, such as proof-of-stake or delegated proof-of-team sprint, could further enhance its applicability across different blockchain architectures.

As blockchain ecosystems evolve, ensuring equitable reward distribution and energy efficiency will remain critical. PoTS presents a compelling framework to achieve these objectives, laying the foundation for a more sustainable and inclusive decentralized economy. Future innovations in team-based consensus mechanisms may further refine these principles, ultimately contributing to the broader advancement of distributed ledger technologies.

\section*{Acknowledgments}
The author declares no specific funding for this research and wishes to acknowledge his institution for providing a supportive research environment.

\bibliographystyle{IEEEtran}
\bibliography{pots_sim}
\end{document}